\documentclass[12pt,a4paper,oneside]{article}
\usepackage{graphicx}
\usepackage{epstopdf}
\usepackage{epsfig}
\usepackage{indentfirst}
\usepackage{authblk}
\usepackage{xcolor}

\title{Determinations of quark mixing matrix elements $|V_{cd}|$ and $|V_{cs}|$ from leptonic and semileptonic $D$ Decays}
\author[1]{G. Rong}
\author[1]{Y. Fang}
\author[1]{H. L. Ma}
\affil[1]{\it \small Institute of High Energy Physics, Beijing 100049, People's Republic of China}

\date{29 August, 2014}

\begin{document}

\maketitle

\begin{abstract}
With the recent measurements of purely leptonic $D^+_{(s)}$ decays and
semileptonic $D$ decays in conjunction with decay constants $f_{D^+_{(s)}}$
and form factors $f^{\pi(K)}_+(0)$ calculated
in LQCD, we extract the magnitudes of $V_{cd}$ and $V_{cs}$ to be
$|V_{cd}|=0.218\pm0.005$ and $|V_{cs}|=0.987\pm0.016$.
Compared to those given in PDG2013, the precisions of these newly extracted $|V_{cd}|$
and $|V_{cs}|$ are improved
by more than 2.0 and 1.5 factors, respectively.
With the newly extracted $|V_{cd}|$ and $|V_{cs}|$ together with other CKM matrix elements
given in PDG2013, we check the unitarity of the CKM matrix, which are
$|V_{ud}|^2+|V_{cd}|^2+|V_{td}|^2=0.997\pm0.002$,
$|V_{us}|^2+|V_{cs}|^2+|V_{ts}|^2=1.027\pm0.032$
and
$|V_{cd}|^2+|V_{cs}|^2+|V_{cb}|^2=1.023\pm0.032$.
\end{abstract}

\section{Introduction}

In the Standard Model (SM) of particle physics, the $D^+_{(s)}$
meson can decay into $\ell^+\nu_\ell$ (where $\ell = e$, $\mu$, or $\tau$)
via annihilation mediated by a virtual $W^+$ boson.
The decay rate depends upon the wave function overlap of
the two quarks at the origin, which is parameterized by the $D^+_{(s)}$ decay
constant, $f_{D^+_{(s)}}$.   All of the strong interaction effects between the two initial-state quarks
are absorbed into $f_{D^+_{(s)}}$.
In the SM, the decay width of $D^+_{(s)}\to\ell^+\nu_\ell$ is given by
\begin{equation}
\Gamma(D^+_{(s)} \rightarrow \ell^+\nu_{\ell})=
     \frac{G^2_F f^2_{D^+_{(s)}}} {8\pi}
     \mid V_{cd(s)} \mid^2
      m^2_{\ell} m_{D^+_{(s)}}
    \left (1- \frac{m^2_{\ell} } {m^2_{D^+_{(s)}}}\right )^2,
\label{eq01}
\end{equation}
where $G_F$ is the Fermi coupling constant,
$V_{cd(s)}$ is the $c\to d(s)$ Cabibbo-Kobayashi-Maskawa (CKM)
matrix element~\cite{pdg2013}, $m_{\ell}$ is the lepton mass, and
$m_{D^+_{(s)}}$ is the $D^+_{(s)}$ meson mass.

    Similarly, in the SM, neglecting the positron mass, the differential decay rate of
$D \to \pi(K) e^+\nu_e$ process
is given by
\begin{equation}
\frac {d\Gamma }{dq^2} = X \frac {G_F^2}{24\pi ^3}|V_{cd(s)}|^2
|\vec p_{\pi(K)}|^3
|f_+^{\pi(K)}(q^2)|^2,
\label{eq_dGamma_dq2}
\end{equation}
where
$\vec p_{\pi(K)}$ is the three-momentum of the
$\pi$ ($K$) meson in the rest frame of the $D$ meson,
$f^{\pi(K)}_+(q^2)$ represents the
hadronic form factor of the hadronic weak current depending on
square of the four-momenta transfer
$q^2$, and $X$ is a factor due to isospin, which equals to $1$ for $D^0\to\pi^-e^+\nu_e$, $D^0\to K^-e^+\nu_e$
and $D^+\to\bar K^0e^+\nu_e$, and equals to $1/2$ for $D^+\to\pi^0e^+\nu_e$.
The form factor $f^{\pi(K)}_+(q^2)$ measures the probability
to form the final state $\pi$ ($K$) meson in this decay.

Recently, the branching fractions for leptonic $D^+$ and $D^+_s$ decays
were well measured at the $e^+e^-$ experiments near threshold of
the $D\bar D$ production (CLEO-c and BESIII)
and near $10.6$ GeV (Belle and BaBar),
and the decay constants $f_{D^+}$ and $f_{D^+_s}$ were calculated in LQCD at
precisions of $\sim 1.6\%$ and $\sim 1.1\%$, respectively.
With these measured branching fractions in conjunction with
the $f_{D^+_{(s)}}$ calculated in LQCD,
the magnitudes of CKM quark mixing parameters $V_{cd}$ and $V_{cs}$ can be well
extracted.
In addition, the precisions of these measured branching fractions
for $D\rightarrow \pi e^+\nu_e$ and $D\rightarrow K e^+\nu_e$ decays
or measured products of $|V_{cd(s)}|$ and $f_+^{\pi(K)}(0)$  are at an accuracy level of
about $1\%$,
while the LQCD calculations of these form factors $f_+^{\pi}(0)$ and $f_+^{K}(0)$ also reach to about
$4.4\%$ and $2.5\%$, respectively.
With these measured products of $|V_{cd(s)}|$ and $f_+^{\pi(K)}(0)$
together with inputs of the form factors calculated in LQCD, the magnitudes of
$V_{cd}$ and $V_{cs}$ can also be well extracted.

In this article, we extract $|V_{cd}|$ and $|V_{cs}|$
with these measured branching fractions and/or $|V_{cd(s)}|f_+^{\pi(K)}(0)$
in conjunction with decay constants $f_{D_{(s)}^+}$ and/or form factors
$f_+^{\pi(K)}(0)$ calculated in LQCD.
In determinations of $|V_{cd}|$ and $|V_{cs}|$,
we use $G_F$, masses of $D^+_{(s)}$ meson and leptons,
and lifetimes of $D^+_{(s)}$ meson given in PDG2013~\cite{pdg2013}.

\section{Recent experimental measurements}

\subsection{Purely leptonic $D^+$ decays}

In 2008, the CLEO-c Collaboration accumulated $460055\pm 787$ $D^-$ tags
by analyzing 818 pb$^{-1}$ data taken at 3.773 GeV and selecting $D^-$ mesons from
6 hadronic decay modes
of the $D^-$ meson.
They observed $149.7 \pm 12.0$ signal events
for $D^+ \rightarrow \mu^+\nu_\mu$ decays in the system recoiling against these $D^-$ tags.
They measured the decay branching fraction
$B(D^+ \rightarrow \mu^+\nu_\mu)=(3.82 \pm 0.32 \pm 0.09)\times 10^{-4}$
~\cite{cleo-c_fD_2008}.

In 2014, the BESIII Collaboration measured the branching fraction
for $D^+ \rightarrow \mu^+\nu_\mu$ decays by analyzing 2.92 fb$^{-1}$ data taken at 3.773 GeV.
From 9 hadronic decay modes of $D^-$ meson, the BESIII Collaboration accumulated $1703054\pm3405$ $D^-$ tags.
In this $D^-$ tag sample they observed $409.0\pm 21.2$ signal events for
$D^+ \rightarrow \mu^+\nu_\mu$ decays and measured branching fraction
$B(D^+ \rightarrow \mu^+\nu_\mu)=(3.71 \pm 0.19 \pm 0.06)\times 10^{-4}$~\cite{BESIII_Dptomunu}.

Averaging these two branching fractions, we obtain
\begin{equation}\label{eq:bf_Dptomunu}
B(D^+\to\mu^+\nu_\mu) = (3.74\pm0.17)\times10^{-4},
\end{equation}
where the error is the combined statistical and systematic errors together.

\subsection{Purely leptonic $D_s^+$ decays}
In 2009, the CLEO-c Collaboration studied the $D_s^+\rightarrow \ell^+\nu_\ell$ decays
based on 600 pb$^{-1}$ data taken at 4.17 GeV.
From this data sample, they tagged $D_s^-$ mesons
from 9 hadronic decay modes.
By examining
distribution of missing mass-squared
of the $D_s^{-}$ and $\gamma$ system
they accumulated $43859\pm 936$ $D_s^{+}$ mesons;
by analyzing distribution of missing mass-squared
of the $D^{-}_s\gamma\mu^+$ system,
they selected $D^+_s\to\mu^+\nu_\mu$ decay events and measured the branching fraction
$B(D^+_s\to\mu^+\nu_\mu)=(0.565\pm0.045\pm0.017)\%$~\cite{CLEOc_Dpstomunu}.
Using similar method, the CLEO-c Collaboration also measured the branching fraction
$B(D^+_s\to\tau^+\nu_\tau) = (5.58\pm0.33\pm0.13)\%$,
which is the average of three measured branching fractions obtained with
$\tau^+\to\pi^+\bar\nu_\tau$~\cite{CLEOc_Dpstomunu}, $\tau^+\to e^+\nu_e\bar\nu_\tau$~\cite{CLEOc_Dpstotau_e} and $\tau^+\to\rho^+\bar\nu_\tau$ decays~\cite{CLEOc_Dpstotau_rho}.

In 2013, the Belle Collaboration measured the branching fractions
for leptonic $D_s^+$ decays.
They selected leptonic $D_s^+$ decays from the $e^+e^- \rightarrow c\bar c$ continuum production,
in which the $D_{\rm tag}K_{\rm frag}X_{\rm frag}D^{*+}_s$ is produced from the quark fragmentation,
where $D_s^{*+} \rightarrow \gamma D^+_s$,
$K_{\rm frag}$ is either $K^+$ or $K^0_S$,
and $X_{\rm frag}$ indicates several pions or photons.
By reconstructing the recoil mass of the $D_{\rm tag}K_{\rm frag}X_{\rm frag} \gamma$, they observed clear
$D^+_s$ signal in the system recoiling against the $D_{\rm tag}K_{\rm frag}X_{\rm frag} \gamma$.
By fitting the recoil mass spectra of $D_{\rm tag}K_{\rm frag}X_{\rm frag} \gamma$,
they accumulated $94360\pm1310\pm1450$ inclusive $D^+_s$ mesons.
To search for $D^+_s\to\mu^+\nu_\mu$ decays, they examined the  missing mass-squared
$M^2_{\rm miss}(D_{\rm tag}K_{\rm frag}X_{\rm frag}\gamma\mu)$
distribution of
the $D_{\rm tag}K_{\rm frag}X_{\rm frag} \gamma \mu$ system.
Fitting the $M^2_{\rm miss}(D_{\rm tag}K_{\rm frag}X_{\rm frag} \gamma \mu)$ distribution yields $492\pm26$
signal events for $D_s^{+} \rightarrow \mu^+ \nu_\mu$ decays.
With these numbers of events,
the Belle Collaboration measured the decay branching fraction
$B(D_s^+ \rightarrow \mu^+\nu)=(0.531 \pm 0.028 \pm 0.020)\%$~\cite{Belle_Dstolnu}.
In addition, the Belle Collaboration observed
$2217 \pm 83$ signal events for $D_s^+ \rightarrow \tau^+\nu_\tau$ decays
with $\tau^+\rightarrow e^+\nu_e\bar\nu_\tau$,
$\tau^+\rightarrow \mu^+\nu_\mu\bar\nu_\tau$ and  $\tau^+\rightarrow \pi^+\bar\nu_\tau$ decays,
and measured
the decay branching fraction
$B(D_s^+ \rightarrow \tau^+\nu_\tau)=(5.70 \pm 0.21^{+0.31}_{-0.30})\%$~\cite{Belle_Dstolnu}.

   In 2010, using the similar technique as the one used by the Belle Collaboration,
the BaBar Collaboration made measurements of the branching fractions for leptonic $D^+_s$ decays.
By analyzing 521 fb$^{-1}$  data taken at 10.6 GeV,
the BaBar Collaboration measured the decay branching fractions
$B(D_s^+ \rightarrow \mu^+\nu_\mu)=(0.602 \pm 0.038 \pm 0.034)\%$
and
$B(D_s^+ \rightarrow \tau^+\nu_\tau)=(5.00 \pm 0.35 \pm 0.49)\%$~\cite{BaBar_Dstolnu}.

Combining these branching fractions measured by the CLEO-c, Belle and BaBar Collaborations,
we obtain
\begin{equation}
 B(D_s^+ \rightarrow \mu^+\nu_\mu)=(0.556 \pm 0.025)\%
\end{equation}
and
\begin{equation}
 B(D_s^+ \rightarrow \tau^+\nu_\tau)=(5.54 \pm 0.24)\%,
\end{equation}
where the errors are the combined statistical and systematic errors together.

\subsection{Semileptonic $D$ decays}

In 2008, the CLEO-c Collaboration studied the semileptonic decays of $D^0\to \pi^-e^+\nu_e$, $D^0\to K^-e^+\nu_e$,
$D^+\to\pi^0e^+\nu_e$ and $D^+\to\bar K^0e^+\nu_e$ by analyzing 818 pb$^{-1}$ data taken at 3.773 GeV.
They extracted the products $f_+^\pi(0)|V_{cd}|=0.150\pm0.004\pm0.001$ and
$f_+^K(0)|V_{cs}|=0.719\pm0.006\pm0.005$
by fitting their measured partial decay rates
with form factor parameterized with three parameter series expansion~\cite{CLEOc_DtoPenu}.

Recently, the BESIII Collaboration reported their new preliminary results of
$D^0 \rightarrow \pi^-e^+\nu_e$ and $D^0\to K^-e^+\nu_e$ decays obtained by analyzing
2.92 fb$^{-1}$ data taken at 3.773 GeV.
They obtained $f_+^\pi(0)|V_{cd}|=0.1420\pm0.0024\pm0.0010$ and
$f_+^K(0)|V_{cs}|=0.7196\pm0.0035\pm0.0041$
by fitting differential decay rates with
the three parameter series expansion~\cite{BESIII_D0toPlnu}.

In 2007, the BaBar Collaboration measured the form factors $f^K_+(q^2)$ by analyzing 75 fb$^{-1}$ data collected
at 10.6 GeV and  determined $f^K_+(0)=0.727\pm0.007\pm0.005\pm0.007$~\cite{BaBar_D0toKenu}.
Multiplying this form factor by $|V_{cs}|=0.9729\pm0.0003$ used in
their paper, we obtain the product $f_+^K(0)|V_{cs}|=0.707\pm0.007\pm0.005\pm0.007$.
Using the same technique, the BaBar Collaboration also studied the $D^0\to\pi^-e^+\nu_e$ decay
by analyzing 347.2 fb$^{-1}$ data collected at $\Upsilon(4S)$
and reported preliminary results at ICHEP2014.
They measured $f^\pi_+(0)|V_{cd}|=0.1374\pm0.0038\pm0.0022\pm0.0009$~\cite{BaBar_D0topienu}.

Combining these $f_+^{\pi(K)}(0)|V_{cd(s)}|$
measured at the CLEO-c, BESIII and BaBar experiments, we obtain
\begin{equation}
f^\pi_+(0)|V_{cd}| = 0.143\pm0.002
\end{equation}
and
\begin{equation}
f^K_+(0)|V_{cs}| = 0.718\pm0.004,
\end{equation}
where the errors are the combined statistical and systematic errors together.

\section{Determinations of $|V_{cd}|$}

Before 2012, the CKM matrix element $|V_{cd}|$ was usually determined with the
$\nu\bar\nu$ interaction or the semileptonic decay of $D\to\pi e^+\nu_e$.
Actually,
using the measured branching fraction for $D^+\to\mu^+\nu_\mu$ decays
in conjunction with the LQCD calculation on  $D^+$ meson decay constant,
the magnitude of $V_{cd}$ can also be extracted via the Eq.~(\ref{eq01}).
At Charm2012, the BESIII Collaboration reported preliminary result on the determination of $|V_{cd}|$
based on their measured branching fraction for
$D^+\to\mu^+\nu_\mu$ decay, which is $|V_{cd}|=0.2218\pm0.0062\pm0.0047$~\cite{Charm2012_Dptomu}.
Recently, the Flavor Lattice Averaging Group (FLAG) made an average of several values of the
$f_{D^+}$ calculated in LQCD. The averaged $D^+$ decay constant calculated in LQCD is
$f_{D^+}=(209.2\pm3.3)$ MeV~\cite{FLAG}.
Inserting the averaged branching fraction for $D^+\to\mu^+\nu_\mu$ decays
as given in Eq.~(\ref{eq:bf_Dptomunu}) and
this averaged $f_{D^+}$ into Eq.~(\ref{eq01}) yields
\begin{equation}\label{eq:Vcd_Dptomu}
|V_{cd}|_{D^+\to\mu^+\nu_\mu} = 0.219\pm0.005\pm0.004,
\end{equation}
where the first uncertainty is from the measured branching fractions
and the second mainly from the uncertainties of
$f_{D^+}$ and the lifetime of $D^+$ meson.

Dividing the averaged $f^\pi_+(0)|V_{cd}|$ from semileptonic $D\to\pi e^+\nu_e$ decays
by the form factor $f_+^\pi(0)=0.666\pm0.029$ calculated in LQCD~\cite{LQCD_fpi}
yields
\begin{equation}
|V_{cd}|_{D\to\pi e^+\nu_e} = 0.215\pm0.003\pm0.009,
\end{equation}
where the first uncertainty is from the measured $f^\pi_+(0)|V_{cd}|$,
and the second uncertainty is from $f^\pi_+(0)$.

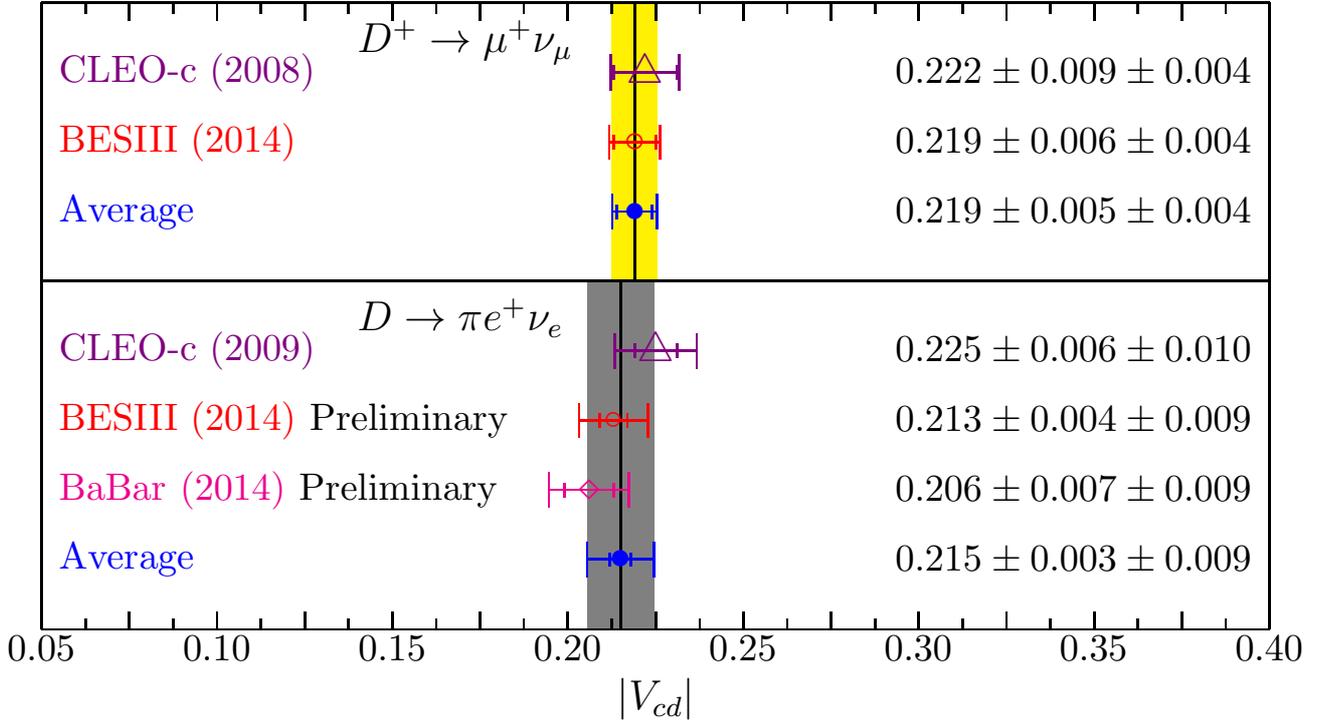
\begin{figure}[!htp]
\setlength{\unitlength}{1pt}
\begin{center}
\resizebox{\textwidth}{!}{
\begin{picture}(350,210)

 \setlength\fboxsep{0pt}
 \put(162.6,120){\colorbox{yellow}{\makebox(12.8, 80){}}}
 \put(169.0,120){\line(0,1){80}}
 \put(155.5, 20){\colorbox{gray}{\makebox(19.0, 100){}}}
 \put(165.0, 20){\line(0,1){100}}
 \put(0, 120){\line(1,0){350}}
 \put(  0, 20){\line(1,0){350}}
 \put(  0,200){\line(1,0){350}}
 \put(  0, 20){\line(0,1){180}}
 \put(350, 20){\line(0,1){180}}
 \multiput(0, 20)(25.0, 0){14}{\line(0, 1){5}}
 \multiput(0, 20)(12.5, 0){28}{\line(0, 1){3}}
 \multiput(0,200)(25.0, 0){14}{\line(0,-1){5}}
 \multiput(0,200)(12.5, 0){28}{\line(0,-1){3}}
 \put(  0, 18){\makebox(0,0)[t]{\small 0.05}}
 \put( 50, 18){\makebox(0,0)[t]{\small 0.10}}
 \put(100, 18){\makebox(0,0)[t]{\small 0.15}}
 \put(150, 18){\makebox(0,0)[t]{\small 0.20}}
 \put(200, 18){\makebox(0,0)[t]{\small 0.25}}
 \put(250, 18){\makebox(0,0)[t]{\small 0.30}}
 \put(300, 18){\makebox(0,0)[t]{\small 0.35}}
 \put(350, 18){\makebox(0,0)[t]{\small 0.40}}
 \put(175,  0){\makebox(0,0)[c]{$|V_{cd}|$}}

\color{violet}
\put(5, 180){\makebox(0,0)[l]{\small CLEO-c (2008)}}
\put(172.0,180.0){\makebox(0,0){$\triangle$}}
\put(172.0,180.0){\line( 1, 0){9.8}}
\put(172.0,180.0){\line(-1, 0){9.8}}
\put(181.0,180.0){\line( 0, 1){2.0}}
\put(181.0,180.0){\line( 0,-1){2.0}}
\put(163.0,180.0){\line( 0, 1){2.0}}
\put(163.0,180.0){\line( 0,-1){2.0}}
\put(181.8,180.0){\line( 0, 1){5.0}}
\put(181.8,180.0){\line( 0,-1){5.0}}
\put(162.2,180.0){\line( 0, 1){5.0}}
\put(162.2,180.0){\line( 0,-1){5.0}}
%
\color{red}
\put(5, 160){\makebox(0,0)[l]{\small BESIII (2014)}}
\put(169.0,160.0){\makebox(0,0){$\circ$}}
\put(169.0,160.0){\line( 1, 0){7.2}}
\put(169.0,160.0){\line(-1, 0){7.2}}
\put(175.0,160.0){\line( 0, 1){2.0}}
\put(175.0,160.0){\line( 0,-1){2.0}}
\put(163.0,160.0){\line( 0, 1){2.0}}
\put(163.0,160.0){\line( 0,-1){2.0}}
\put(176.2,160.0){\line( 0, 1){5.0}}
\put(176.2,160.0){\line( 0,-1){5.0}}
\put(161.8,160.0){\line( 0, 1){5.0}}
\put(161.8,160.0){\line( 0,-1){5.0}}

\color{blue}
\put(5, 140){\makebox(0,0)[l]{\small Average}}
\put(169.0,140.0){\makebox(0,0){$\bullet$}}
\put(169.0,140.0){\line( 1, 0){6.4}}
\put(169.0,140.0){\line(-1, 0){6.4}}
\put(174.0,140.0){\line( 0, 1){2.0}}
\put(174.0,140.0){\line( 0,-1){2.0}}
\put(164.0,140.0){\line( 0, 1){2.0}}
\put(164.0,140.0){\line( 0,-1){2.0}}
\put(175.4,140.0){\line( 0, 1){5.0}}
\put(175.4,140.0){\line( 0,-1){5.0}}
\put(162.6,140.0){\line( 0, 1){5.0}}
\put(162.6,140.0){\line( 0,-1){5.0}}

\color{violet}
\put(5, 100){\makebox(0,0)[l]{\small CLEO-c (2009)}}
\put(175.0,100.0){\makebox(0,0){$\triangle$}}
\put(175.0,100.0){\line( 1, 0){11.7}}
\put(175.0,100.0){\line(-1, 0){11.7}}
\put(181.0,100.0){\line( 0, 1){2.0}}
\put(181.0,100.0){\line( 0,-1){2.0}}
\put(169.0,100.0){\line( 0, 1){2.0}}
\put(169.0,100.0){\line( 0,-1){2.0}}
\put(186.7,100.0){\line( 0, 1){5.0}}
\put(186.7,100.0){\line( 0,-1){5.0}}
\put(163.3,100.0){\line( 0, 1){5.0}}
\put(163.3,100.0){\line( 0,-1){5.0}}

\color{red}
\put(5, 80){\makebox(0,0)[l]{\small BESIII (2014) \color{black} Preliminary}}
\put(163.0,80.0){\makebox(0,0){\small $\circ$}}
\put(163.0,80.0){\line( 1, 0){9.8}}
\put(163.0,80.0){\line(-1, 0){9.8}}
\put(167.0,80.0){\line( 0, 1){2.0}}
\put(167.0,80.0){\line( 0,-1){2.0}}
\put(159.0,80.0){\line( 0, 1){2.0}}
\put(159.0,80.0){\line( 0,-1){2.0}}
\put(172.8,80.0){\line( 0, 1){5.0}}
\put(172.8,80.0){\line( 0,-1){5.0}}
\put(153.2,80.0){\line( 0, 1){5.0}}
\put(153.2,80.0){\line( 0,-1){5.0}}

\color{magenta}
\put(5, 60){\makebox(0,0)[l]{\small BaBar (2014) \color{black} Preliminary}}
\put(156.0,60.0){\makebox(0,0){$\diamond$}}
\put(156.0,60.0){\line( 1, 0){11.4}}
\put(156.0,60.0){\line(-1, 0){11.4}}
\put(163.0,60.0){\line( 0, 1){2.0}}
\put(163.0,60.0){\line( 0,-1){2.0}}
\put(149.0,60.0){\line( 0, 1){2.0}}
\put(149.0,60.0){\line( 0,-1){2.0}}
\put(167.4,60.0){\line( 0, 1){5.0}}
\put(167.4,60.0){\line( 0,-1){5.0}}
\put(144.6,60.0){\line( 0, 1){5.0}}
\put(144.6,60.0){\line( 0,-1){5.0}}

\color{blue}
\put(5, 40){\makebox(0,0)[l]{\small Average}}
\put(165.0,40.0){\makebox(0,0){$\bullet$}}
\put(165.0,40.0){\line( 1, 0){9.5}}
\put(165.0,40.0){\line(-1, 0){9.5}}
\put(168.0,40.0){\line( 0, 1){2.0}}
\put(168.0,40.0){\line( 0,-1){2.0}}
\put(162.0,40.0){\line( 0, 1){2.0}}
\put(162.0,40.0){\line( 0,-1){2.0}}
\put(174.5,40.0){\line( 0, 1){5.0}}
\put(174.5,40.0){\line( 0,-1){5.0}}
\put(155.5,40.0){\line( 0, 1){5.0}}
\put(155.5,40.0){\line( 0,-1){5.0}}

\color{black}
\put(90, 195){\makebox(0,0)[lt]{$D^+\to\mu^+\nu_\mu$}}
\put(90, 115){\makebox(0,0)[lt]{$D\to\pi e^+\nu_e$}}

\put(345, 180){\makebox(0,0)[r]{\small $0.222\pm0.009\pm0.004$}}
\put(345, 160){\makebox(0,0)[r]{\small $0.219\pm0.006\pm0.004$}}
\put(345, 140){\makebox(0,0)[r]{\small $0.219\pm0.005\pm0.004$}}
\put(345, 100){\makebox(0,0)[r]{\small $0.225\pm0.006\pm0.010$}}
\put(345,  80){\makebox(0,0)[r]{\small $0.213\pm0.004\pm0.009$}}
\put(345,  60){\makebox(0,0)[r]{\small $0.206\pm0.007\pm0.009$}}
\put(345,  40){\makebox(0,0)[r]{\small $0.215\pm0.003\pm0.009$}}
\end{picture}
}
\end{center}
\caption{
Comparison of $|V_{cd}|$ determined from leptonic $D^+$ and semileptonic $D$ decays.
}
\label{fig:Vcd}
\end{figure}

\begin{figure}[!htp]
\setlength{\unitlength}{1pt}
\begin{center}
\resizebox{\textwidth}{!}{
\begin{picture}(350,140)

 \put(100, 20){\line(1,0){150}}
 \put(100,140){\line(1,0){150}}
 \put(100, 20){\line(0,1){120}}
 \put(250, 20){\line(0,1){120}}
 \multiput(100, 20)(30.0, 0){ 5}{\line(0, 1){5}}
 \multiput(100, 20)(15.0, 0){10}{\line(0, 1){3}}
 \multiput(100,140)(30.0, 0){ 5}{\line(0,-1){5}}
 \multiput(100,140)(15.0, 0){10}{\line(0,-1){3}}
 \put(100, 18){\makebox(0,0)[t]{\small 0.20}}
 \put(130, 18){\makebox(0,0)[t]{\small 0.21}}
 \put(160, 18){\makebox(0,0)[t]{\small 0.22}}
 \put(190, 18){\makebox(0,0)[t]{\small 0.23}}
 \put(220, 18){\makebox(0,0)[t]{\small 0.24}}
 \put(250, 18){\makebox(0,0)[t]{\small 0.25}}
 \put(175,  0){\makebox(0,0)[c]{$|V_{cd}|$}}

\color{blue}
\put(10, 100){\makebox(0,0)[l]{PDG2013}}
\put(190.0,100.0){\makebox(0,0){$\triangle$}}
\put(190.0,100.0){\line( 1, 0){33.0}}
\put(190.0,100.0){\line(-1, 0){33.0}}
\put(223.0,100.0){\line( 0, 1){10.0}}
\put(223.0,100.0){\line( 0,-1){10.0}}
\put(157.0,100.0){\line( 0, 1){10.0}}
\put(157.0,100.0){\line( 0,-1){10.0}}
\color{red}
\put(10, 60){\makebox(0,0)[l]{This work}}
\put(154.0,60.0){\makebox(0,0){$\bigcirc$}}
\put(154.0,60.0){\line( 1, 0){15.0}}
\put(154.0,60.0){\line(-1, 0){15.0}}
\put(169.0,60.0){\line( 0, 1){10.0}}
\put(169.0,60.0){\line( 0,-1){10.0}}
\put(139.0,60.0){\line( 0, 1){10.0}}
\put(139.0,60.0){\line( 0,-1){10.0}}
\put(340, 100){\makebox(0,0)[r]{\color{black} $0.230\pm0.011$}}
\put(340,  60){\makebox(0,0)[r]{\color{black} $0.218\pm0.005$}}
\end{picture}
}
\end{center}
\caption{
Comparison of the newly determined $|V_{cd}|$ from
both the leptonic $D^+$ and semileptonic $D$ decays and the one given in PDG2013.
}
\label{fig:Cmp_Vcd_PDG2013}
\end{figure}
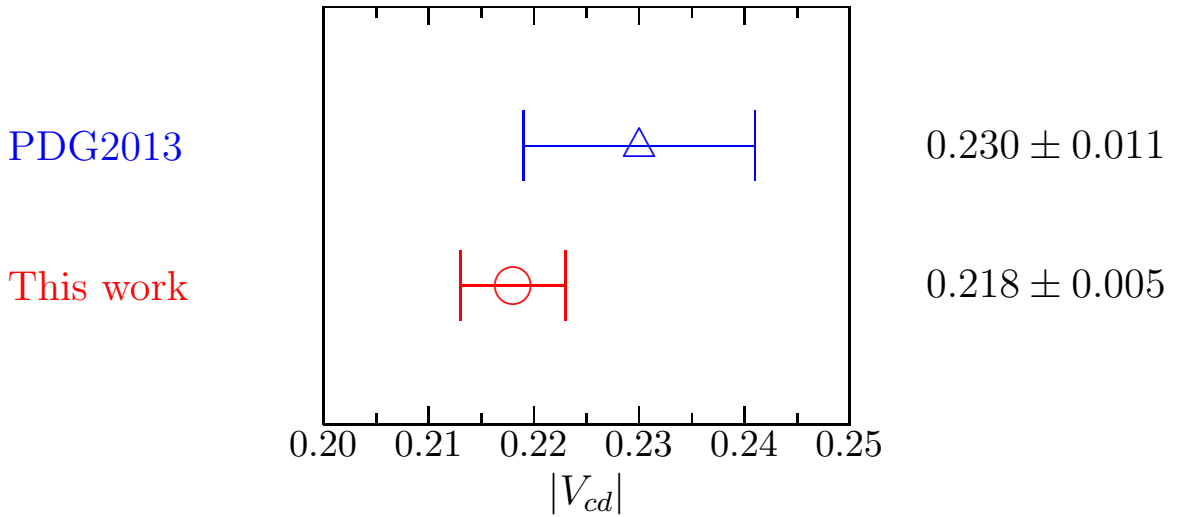

Figure~\ref{fig:Vcd} shows the comparison of $|V_{cd}|$ determined from purely
leptonic $D^+$ decay and semileptonic $D$ decay.
Averaging the determined $|V_{cd}|_{D^+\to\mu^+\nu_\mu}$ and
$|V_{cd}|_{D\to\pi e^+\nu_e}$
yields
\begin{equation}\label{eq:Vcd_Leptonic}
|V_{cd}| = 0.218\pm0.005.
\end{equation}
Figure~\ref{fig:Cmp_Vcd_PDG2013} shows the comparison of
the newly determined $|V_{cd}|$ and
the one given in PDG2013~\cite{pdg2013}.

\section{Determinations of $|V_{cs}|$}

Using the measured decay branching fractions for $D^+_s\to\ell^+\nu_\ell$
together with
the $D^+_s$ meson decay constant calculated in LQCD,
the magnitude of $V_{cs}$ can be extracted via Eq.~(\ref{eq01}).
We herein use the value of $f_{D^+_s}=(248.6\pm2.7)~\rm MeV$,
which is the FLAG average of several decay constants calculated in LQCD~\cite{FLAG},
to extract $|V_{cs}|$.
Inserting the averaged branching fractions for $D^+_s\to\ell^+\nu_\ell$ decays
and the $f_{D^+_s}$ into Eq.~(\ref{eq01}) yields
\begin{equation}\label{eq:Vcd_Dpstomu}
|V_{cs}|_{D^+_s\to\mu^+\nu_\mu} = 1.001\pm0.022\pm0.013
\end{equation}
and
\begin{equation}\label{eq:Vcd_Dpstotau}
|V_{cs}|_{D^+_s\to\tau^+\nu_\tau} = 1.011\pm0.022\pm0.013,
\end{equation}
where the first uncertainties are from the measured branching fractions,
and
the second uncertainties are mainly from $f_{D^+_s}$ and the lifetime of $D^+_s$ meson.
Combining the above two values, we obtain
\begin{equation}\label{eq:Vcd_Dpstomu}
|V_{cs}|_{D^+_s\to\ell^+\nu_\ell} = 1.006\pm0.016\pm0.013,
\end{equation}
where the first uncertainty is from the measured branching fractions,
the second uncertainty is mainly from $f_{D^+_s}$ and the lifetime of $D^+_s$ meson.

Dividing the averaged $f^K_+(0)|V_{cs}|$ from semileptonic $D\to K e^+\nu_e$ decays
by the form factor
$f_+^K(0)=0.747\pm0.019$ calculated in LQCD~\cite{LQCD_fK} yields
\begin{equation}
|V_{cs}|_{D\to K e^+\nu_e} = 0.961\pm0.005\pm0.024,
\end{equation}
where the first uncertainty is from the measured $f^K_+(0)|V_{cs}|$
and the second uncertainty is from $f^K_+(0)$.

\begin{figure}[!htp]
\setlength{\unitlength}{1pt}
\begin{center}
\resizebox{\textwidth}{!}{
\begin{picture}(350,320)
 \setlength\fboxsep{0pt}
 \put(162.7,220){\colorbox{green}{\makebox(25.6, 100){}}}
 \put(175.5,220){\line(0,1){100}}
 \put(167.7,120){\colorbox{yellow}{\makebox(25.6, 100){}}}
 \put(180.5,120){\line(0,1){100}}
 \put(143.2, 20){\colorbox{gray}{\makebox(24.6, 100){}}}
 \put(155.5, 20){\line(0,1){100}}
 \put(0, 120){\line(1,0){350}}
 \put(0, 220){\line(1,0){350}}
 \put(  0, 20){\line(1,0){350}}
 \put(  0,320){\line(1,0){350}}
 \put(  0, 20){\line(0,1){300}}
 \put(350, 20){\line(0,1){300}}
 \multiput(0, 20)(25.0, 0){14}{\line(0, 1){5}}
 \multiput(0, 20)(12.5, 0){28}{\line(0, 1){3}}
 \multiput(0,320)(25.0, 0){14}{\line(0,-1){5}}
 \multiput(0,320)(12.5, 0){28}{\line(0,-1){3}}
 \put( 25, 18){\makebox(0,0)[t]{\small 0.7}}
 \put( 75, 18){\makebox(0,0)[t]{\small 0.8}}
 \put(125, 18){\makebox(0,0)[t]{\small 0.9}}
 \put(175, 18){\makebox(0,0)[t]{\small 1.0}}
 \put(225, 18){\makebox(0,0)[t]{\small 1.1}}
 \put(275, 18){\makebox(0,0)[t]{\small 1.2}}
 \put(325, 18){\makebox(0,0)[t]{\small 1.3}}
 \put(175,  0){\makebox(0,0)[c]{$|V_{cs}|$}}

\color{violet}
\put(5, 300){\makebox(0,0)[l]{\small CLEO-c (2009)}}
\put(179.5,300.0){\makebox(0,0){$\triangle$}}
\put(179.5,300.0){\line( 1, 0){22.4}}
\put(179.5,300.0){\line(-1, 0){22.4}}
\put(199.5,300.0){\line( 0, 1){2.0}}
\put(199.5,300.0){\line( 0,-1){2.0}}
\put(159.5,300.0){\line( 0, 1){2.0}}
\put(159.5,300.0){\line( 0,-1){2.0}}
\put(201.9,300.0){\line( 0, 1){5.0}}
\put(201.9,300.0){\line( 0,-1){5.0}}
\put(157.1,300.0){\line( 0, 1){5.0}}
\put(157.1,300.0){\line( 0,-1){5.0}}
%

\color{cyan}
\put(5, 280){\makebox(0,0)[l]{\small Belle (2013)}}
\put(164.0,280.0){\makebox(0,0){$\bullet$}}
\put(164.0,280.0){\line( 1, 0){17.0}}
\put(164.0,280.0){\line(-1, 0){17.0}}
\put(177.0,280.0){\line( 0, 1){2.0}}
\put(177.0,280.0){\line( 0,-1){2.0}}
\put(151.0,280.0){\line( 0, 1){2.0}}
\put(151.0,280.0){\line( 0,-1){2.0}}
\put(181.0,280.0){\line( 0, 1){5.0}}
\put(181.0,280.0){\line( 0,-1){5.0}}
\put(147.0,280.0){\line( 0, 1){5.0}}
\put(147.0,280.0){\line( 0,-1){5.0}}

\color{magenta}
\put(5, 260){\makebox(0,0)[l]{\small BaBar (2010)}}
\put(195.5,260.0){\makebox(0,0){$\diamond$}}
\put(195.5,260.0){\line( 1, 0){23.0}}
\put(195.5,260.0){\line(-1, 0){23.0}}
\put(212.0,260.0){\line( 0, 1){2.0}}
\put(212.0,260.0){\line( 0,-1){2.0}}
\put(179.0,260.0){\line( 0, 1){2.0}}
\put(179.0,260.0){\line( 0,-1){2.0}}
\put(218.5,260.0){\line( 0, 1){5.0}}
\put(218.5,260.0){\line( 0,-1){5.0}}
\put(172.5,260.0){\line( 0, 1){5.0}}
\put(172.5,260.0){\line( 0,-1){5.0}}
%

\color{blue}
\put(5, 240){\makebox(0,0)[l]{\small Average}}
\put(175.5,240.0){\makebox(0,0){$\bullet$}}
\put(175.5,240.0){\line( 1, 0){12.8}}
\put(175.5,240.0){\line(-1, 0){12.8}}
\put(186.5,240.0){\line( 0, 1){2.0}}
\put(186.5,240.0){\line( 0,-1){2.0}}
\put(164.5,240.0){\line( 0, 1){2.0}}
\put(164.5,240.0){\line( 0,-1){2.0}}
\put(188.3,240.0){\line( 0, 1){5.0}}
\put(188.3,240.0){\line( 0,-1){5.0}}
\put(162.7,240.0){\line( 0, 1){5.0}}
\put(162.7,240.0){\line( 0,-1){5.0}}

\color{violet}
\put(5, 200){\makebox(0,0)[l]{\small CLEO-c (2009)}}
\put(182.5,200.0){\makebox(0,0){$\triangle$}}
\put(182.5,200.0){\line( 1, 0){17.5}}
\put(182.5,200.0){\line(-1, 0){17.5}}
\put(197.5,200.0){\line( 0, 1){2.0}}
\put(197.5,200.0){\line( 0,-1){2.0}}
\put(167.5,200.0){\line( 0, 1){2.0}}
\put(167.5,200.0){\line( 0,-1){2.0}}
\put(200.0,200.0){\line( 0, 1){5.0}}
\put(200.0,200.0){\line( 0,-1){5.0}}
\put(165.0,200.0){\line( 0, 1){5.0}}
\put(165.0,200.0){\line( 0,-1){5.0}}

\color{cyan}
\put(5, 180){\makebox(0,0)[l]{\small Belle (2013)}}
\put(187.5,180.0){\makebox(0,0){$\bullet$}}
\put(187.5,180.0){\line( 1, 0){18.2}}
\put(187.5,180.0){\line(-1, 0){18.2}}
\put(197.0,180.0){\line( 0, 1){2.0}}
\put(197.0,180.0){\line( 0,-1){2.0}}
\put(178.0,180.0){\line( 0, 1){2.0}}
\put(178.0,180.0){\line( 0,-1){2.0}}
\put(205.7,180.0){\line( 0, 1){5.0}}
\put(205.7,180.0){\line( 0,-1){5.0}}
\put(169.3,180.0){\line( 0, 1){5.0}}
\put(169.3,180.0){\line( 0,-1){5.0}}

\color{magenta}
\put(5, 160){\makebox(0,0)[l]{\small BaBar (2010)}}
\put(155.0,160.0){\makebox(0,0){$\diamond$}}
\put(155.0,160.0){\line( 1, 0){29.8}}
\put(155.0,160.0){\line(-1, 0){29.8}}
\put(172.0,160.0){\line( 0, 1){2.0}}
\put(172.0,160.0){\line( 0,-1){2.0}}
\put(138.0,160.0){\line( 0, 1){2.0}}
\put(138.0,160.0){\line( 0,-1){2.0}}
\put(184.8,160.0){\line( 0, 1){5.0}}
\put(184.8,160.0){\line( 0,-1){5.0}}
\put(125.2,160.0){\line( 0, 1){5.0}}
\put(125.2,160.0){\line( 0,-1){5.0}}

\color{blue}
\put(5, 140){\makebox(0,0)[l]{\small Average}}
\put(180.5,140.0){\makebox(0,0){$\bullet$}}
\put(180.5,140.0){\line( 1, 0){12.8}}
\put(180.5,140.0){\line(-1, 0){12.8}}
\put(191.5,140.0){\line( 0, 1){2.0}}
\put(191.5,140.0){\line( 0,-1){2.0}}
\put(169.5,140.0){\line( 0, 1){2.0}}
\put(169.5,140.0){\line( 0,-1){2.0}}
\put(193.3,140.0){\line( 0, 1){5.0}}
\put(193.3,140.0){\line( 0,-1){5.0}}
\put(167.7,140.0){\line( 0, 1){5.0}}
\put(167.7,140.0){\line( 0,-1){5.0}}

\color{violet}
\put(5, 100){\makebox(0,0)[l]{\small CLEO-c (2009)}}
\put(156.5,100.0){\makebox(0,0){$\triangle$}}
\put(156.5,100.0){\line( 1, 0){13.0}}
\put(156.5,100.0){\line(-1, 0){13.0}}
\put(161.5,100.0){\line( 0, 1){2.0}}
\put(161.5,100.0){\line( 0,-1){2.0}}
\put(151.5,100.0){\line( 0, 1){2.0}}
\put(151.5,100.0){\line( 0,-1){2.0}}
\put(169.5,100.0){\line( 0, 1){5.0}}
\put(169.5,100.0){\line( 0,-1){5.0}}
\put(143.5,100.0){\line( 0, 1){5.0}}
\put(143.5,100.0){\line( 0,-1){5.0}}

\color{red}
\put(5, 80){\makebox(0,0)[l]{\small BESIII (2014) \color{black} Preliminary}}
\put(156.5,80.0){\makebox(0,0){$\circ$}}
\put(156.5,80.0){\line( 1, 0){12.5}}
\put(156.5,80.0){\line(-1, 0){12.5}}
\put(160.0,80.0){\line( 0, 1){2.0}}
\put(160.0,80.0){\line( 0,-1){2.0}}
\put(153.0,80.0){\line( 0, 1){2.0}}
\put(153.0,80.0){\line( 0,-1){2.0}}
\put(169.0,80.0){\line( 0, 1){5.0}}
\put(169.0,80.0){\line( 0,-1){5.0}}
\put(144.0,80.0){\line( 0, 1){5.0}}
\put(144.0,80.0){\line( 0,-1){5.0}}

\color{magenta}
\put(5, 60){\makebox(0,0)[l]{\small BaBar (2007)}}
\put(148.0,60.0){\makebox(0,0){$\diamond$}}
\put(148.0,60.0){\line( 1, 0){14.2}}
\put(148.0,60.0){\line(-1, 0){14.2}}
\put(155.5,60.0){\line( 0, 1){2.0}}
\put(155.5,60.0){\line( 0,-1){2.0}}
\put(140.5,60.0){\line( 0, 1){2.0}}
\put(140.5,60.0){\line( 0,-1){2.0}}
\put(162.2,60.0){\line( 0, 1){5.0}}
\put(162.2,60.0){\line( 0,-1){5.0}}
\put(133.8,60.0){\line( 0, 1){5.0}}
\put(133.8,60.0){\line( 0,-1){5.0}}

\color{blue}
\put(5, 40){\makebox(0,0)[l]{\small Average}}
\put(155.5,40.0){\makebox(0,0){$\bullet$}}
\put(155.5,40.0){\line( 1, 0){12.3}}
\put(155.5,40.0){\line(-1, 0){12.3}}
\put(158.0,40.0){\line( 0, 1){2.0}}
\put(158.0,40.0){\line( 0,-1){2.0}}
\put(153.0,40.0){\line( 0, 1){2.0}}
\put(153.0,40.0){\line( 0,-1){2.0}}
\put(167.8,40.0){\line( 0, 1){5.0}}
\put(167.8,40.0){\line( 0,-1){5.0}}
\put(143.2,40.0){\line( 0, 1){5.0}}
\put(143.2,40.0){\line( 0,-1){5.0}}

\color{black}
\put(75, 315){\makebox(0,0)[lt]{$D^+_s\to\mu^+\nu_\mu$}}
\put(75, 215){\makebox(0,0)[lt]{$D^+_s\to\tau^+\nu_\tau$}}
\put(75, 115){\makebox(0,0)[lt]{$D\to K e^+\nu_e$}}
\put(345, 300){\makebox(0,0)[r]{\small $1.009\pm0.040\pm0.020$}}
\put(345, 280){\makebox(0,0)[r]{\small $0.978\pm0.026\pm0.022$}}
\put(345, 260){\makebox(0,0)[r]{\small $1.041\pm0.033\pm0.032$}}
\put(345, 240){\makebox(0,0)[r]{\small $1.001\pm0.022\pm0.013$}}
\put(345, 200){\makebox(0,0)[r]{\small $1.015\pm0.030\pm0.018$}}
\put(345, 180){\makebox(0,0)[r]{\small $1.025\pm0.019\pm0.031$}}
\put(345, 160){\makebox(0,0)[r]{\small $0.960\pm0.034\pm0.049$}}
\put(345, 140){\makebox(0,0)[r]{\small $1.011\pm0.022\pm0.013$}}
\put(345, 100){\makebox(0,0)[r]{\small $0.963\pm0.010\pm0.024$}}
\put(345,  80){\makebox(0,0)[r]{\small $0.963\pm0.007\pm0.024$}}
\put(345,  60){\makebox(0,0)[r]{\small $0.946\pm0.015\pm0.024$}}
\put(345,  40){\makebox(0,0)[r]{\small $0.961\pm0.005\pm0.024$}}
\end{picture}
}
\end{center}
\caption{
Comparison of $|V_{cs}|$ determined from leptonic $D^+_s$ decays and semileptonic $D\to K e^+\nu_e$ decays.
}
\label{fig:Vcs}
\end{figure}
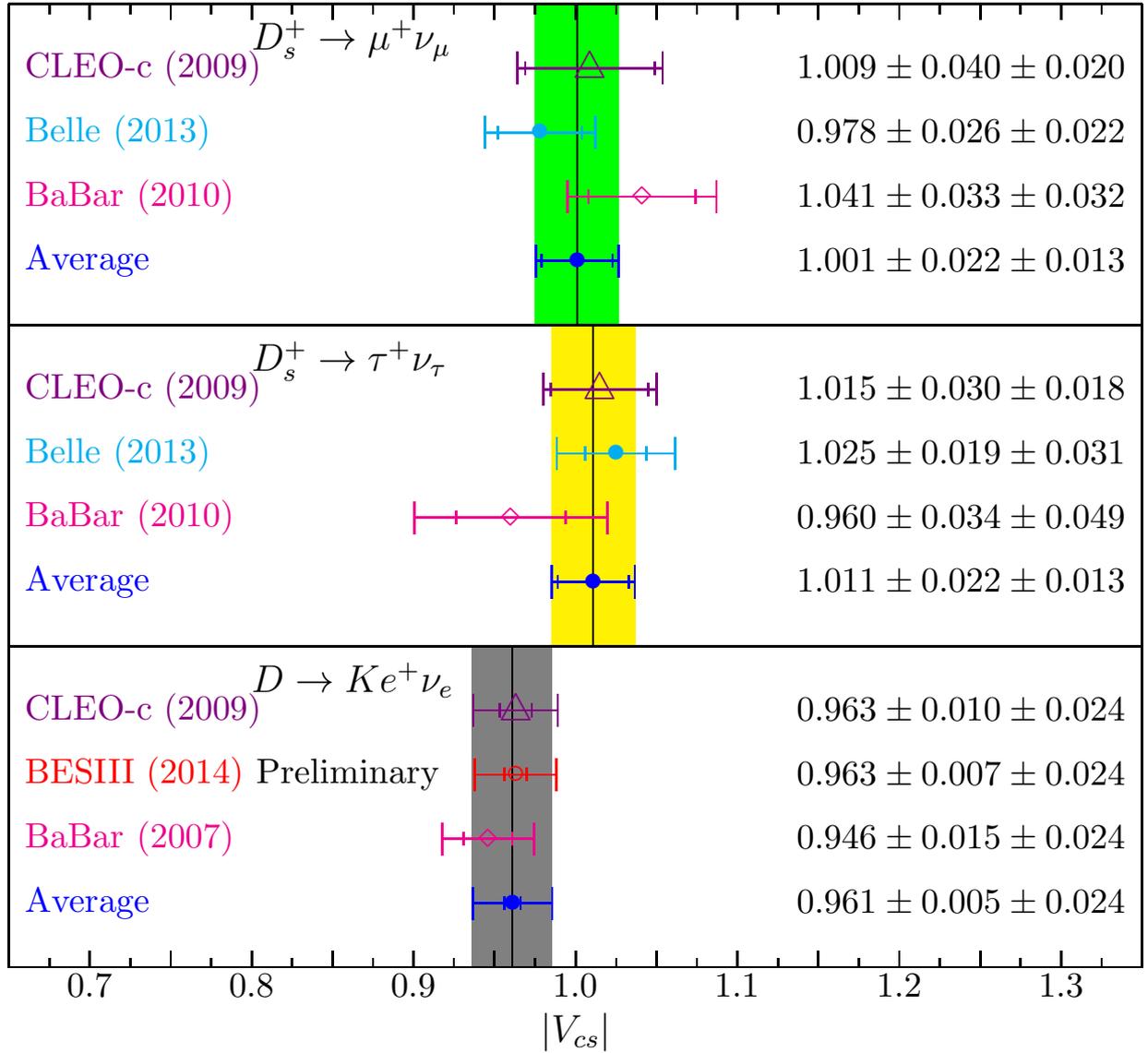

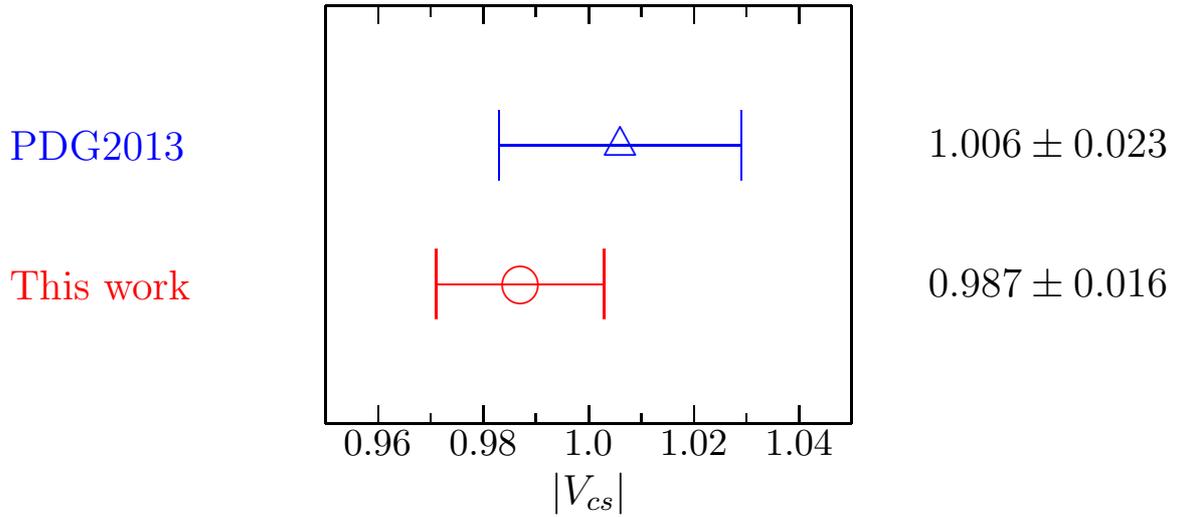
\begin{figure}[!htp]
\setlength{\unitlength}{1pt}
\begin{center}
\resizebox{\textwidth}{!}{
\begin{picture}(350,140)

 \put(100, 20){\line(1,0){150}}
 \put(100,140){\line(1,0){150}}
 \put(100, 20){\line(0,1){120}}
 \put(250, 20){\line(0,1){120}}
 \multiput(115, 20)(30.0, 0){ 5}{\line(0, 1){5}}
 \multiput(115, 20)(15.0, 0){10}{\line(0, 1){3}}
 \multiput(115,140)(30.0, 0){ 5}{\line(0,-1){5}}
 \multiput(115,140)(15.0, 0){10}{\line(0,-1){3}}
 \put(115, 18){\makebox(0,0)[t]{\small 0.96}}
 \put(145, 18){\makebox(0,0)[t]{\small 0.98}}
 \put(175, 18){\makebox(0,0)[t]{\small 1.0}}
 \put(205, 18){\makebox(0,0)[t]{\small 1.02}}
 \put(235, 18){\makebox(0,0)[t]{\small 1.04}}
 \put(175,  0){\makebox(0,0)[c]{$|V_{cs}|$}}

\color{blue}
\put(10, 100){\makebox(0,0)[l]{PDG2013}}
\put(184.0,100.0){\makebox(0,0){$\triangle$}}
\put(184.0,100.0){\line( 1, 0){34.5}}
\put(184.0,100.0){\line(-1, 0){34.5}}
\put(218.5,100.0){\line( 0, 1){10.0}}
\put(218.5,100.0){\line( 0,-1){10.0}}
\put(149.5,100.0){\line( 0, 1){10.0}}
\put(149.5,100.0){\line( 0,-1){10.0}}
\color{red}
\put(10, 60){\makebox(0,0)[l]{This work}}
\put(155.5,60.0){\makebox(0,0){$\bigcirc$}}
\put(155.5,60.0){\line( 1, 0){24.0}}
\put(155.5,60.0){\line(-1, 0){24.0}}
\put(179.5,60.0){\line( 0, 1){10.0}}
\put(179.5,60.0){\line( 0,-1){10.0}}
\put(131.5,60.0){\line( 0, 1){10.0}}
\put(131.5,60.0){\line( 0,-1){10.0}}
\put(340, 100){\makebox(0,0)[r]{\color{black} $1.006\pm0.023$}}
\put(340,  60){\makebox(0,0)[r]{\color{black} $0.987\pm0.016$}}
\end{picture}
}
\end{center}
\caption{
Comparison of the newly determined $|V_{cs}|$ from both the leptonic $D^+_s$ and semileptonic $D$ decays and the one given in PDG2013.
}
\label{fig:Cmp_Vcs_PDG2013}
\end{figure}

Figure~\ref{fig:Vcs} shows the comparison of $|V_{cs}|$ determined from  purely
leptonic $D^+_s$ decays and semileptonic $D$ decays.
Averaging the determined $|V_{cs}|_{D^+_s\to\ell^+\nu_\ell}$ and
$|V_{cs}|_{D\to K e^+\nu_e}$
yields
\begin{equation}\label{eq:Vcs_Leptonic}
|V_{cs}| = 0.987\pm0.016.
\end{equation}
Figure~\ref{fig:Cmp_Vcs_PDG2013} shows the comparison of
the newly determined $|V_{cs}|$ and
the one given in PDG2013~\cite{pdg2013}.

\section{Unitarity checks}
\begin{figure}[!hbp]
\setlength{\unitlength}{1pt}
\begin{center}
\resizebox{\textwidth}{!}{
\begin{picture}(360,200)
 \put(  0, 20){\line(1,0){360}}
 \put(  0,200){\line(1,0){360}}
 \put(  0, 20){\line(0,1){180}}
 \put(360, 20){\line(0,1){180}}
 \multiput(20, 20)(40.0, 0){ 9}{\line(0, 1){5}}
 \multiput(0, 20)(20.0, 0){18}{\line(0, 1){3}}
 \multiput(20,200)(40.0, 0){ 9}{\line(0,-1){5}}
 \multiput(0,200)(20.0, 0){18}{\line(0,-1){3}}
 \put( 20, 18){\makebox(0,0)[t]{\small 0.85}}
 \put( 60, 18){\makebox(0,0)[t]{\small 0.90}}
 \put(100, 18){\makebox(0,0)[t]{\small 0.95}}
 \put(140, 18){\makebox(0,0)[t]{\small 1.0}}
 \put(180, 18){\makebox(0,0)[t]{\small 1.05}}
 \put(220, 18){\makebox(0,0)[t]{\small 1.1}}
 \put(260, 18){\makebox(0,0)[t]{\small 1.15}}
 \put(300, 18){\makebox(0,0)[t]{\small 1.2}}
 \put(340, 18){\makebox(0,0)[t]{\small 1.25}}
\color{red}
\put(140, 20){\line(0,1){180}}

\color{blue}
\put(10, 170){\makebox(0,0)[l]{\small $|V_{ud}|^2+|V_{cd}|^2+|V_{td}|^2$}}
\put(137.6,180.0){\makebox(0,0){\small $\circ$}}
\put(137.6,180.0){\line( 1, 0){1.6}}
\put(137.6,180.0){\line(-1, 0){1.6}}
\put(139.2,180.0){\line( 0, 1){5.0}}
\put(139.2,180.0){\line( 0,-1){5.0}}
\put(136.0,180.0){\line( 0, 1){5.0}}
\put(136.0,180.0){\line( 0,-1){5.0}}
\color{black}
\put(141.6,160.0){\makebox(0,0){$\circ$}}
\put(141.6,160.0){\line( 1, 0){4.0}}
\put(141.6,160.0){\line(-1, 0){4.0}}
\put(145.6,160.0){\line( 0, 1){5.0}}
\put(145.6,160.0){\line( 0,-1){5.0}}
\put(137.6,160.0){\line( 0, 1){5.0}}
\put(137.6,160.0){\line( 0,-1){5.0}}
%
\color{green}
\put(10, 110){\makebox(0,0)[l]{\small $|V_{us}|^2+|V_{cs}|^2+|V_{ts}|^2$}}
\put(161.6,120.0){\makebox(0,0){$\diamond$}}
\put(161.6,120.0){\line( 1, 0){25.6}}
\put(161.6,120.0){\line(-1, 0){25.6}}
\put(187.2,120.0){\line( 0, 1){2.0}}
\put(187.2,120.0){\line( 0,-1){2.0}}
\put(136.0,120.0){\line( 0, 1){2.0}}
\put(136.0,120.0){\line( 0,-1){2.0}}
\put(187.2,120.0){\line( 0, 1){5.0}}
\put(187.2,120.0){\line( 0,-1){5.0}}
\put(136.0,120.0){\line( 0, 1){5.0}}
\put(136.0,120.0){\line( 0,-1){5.0}}

\color{black}
\put(192.0,100.0){\makebox(0,0){$\diamond$}}
\put(192.0,100.0){\line( 1, 0){36.8}}
\put(192.0,100.0){\line(-1, 0){36.8}}
\put(228.8,100.0){\line( 0, 1){5.0}}
\put(228.8,100.0){\line( 0,-1){5.0}}
\put(155.2,100.0){\line( 0, 1){5.0}}
\put(155.2,100.0){\line( 0,-1){5.0}}
%

\color{violet}
\put(10, 50){\makebox(0,0)[l]{\small $|V_{cd}|^2+|V_{cs}|^2+|V_{cb}|^2$}}
\put(158.4,60.0){\makebox(0,0){$\bullet$}}
\put(158.4,60.0){\line( 1, 0){25.6}}
\put(158.4,60.0){\line(-1, 0){25.6}}
\put(184.0,60.0){\line( 0, 1){5.0}}
\put(184.0,60.0){\line( 0,-1){5.0}}
\put(132.8,60.0){\line( 0, 1){5.0}}
\put(132.8,60.0){\line( 0,-1){5.0}}
\color{black}
\put(193.6,40.0){\makebox(0,0){$\bullet$}}
\put(193.6,40.0){\line( 1, 0){37.6}}
\put(193.6,40.0){\line(-1, 0){37.6}}
\put(231.2,40.0){\line( 0, 1){5.0}}
\put(231.2,40.0){\line( 0,-1){5.0}}
\put(156.0,40.0){\line( 0, 1){5.0}}
\put(156.0,40.0){\line( 0,-1){5.0}}
\color{blue}
\put(235, 180){\makebox(0,0)[l]{\small $0.997\pm0.002$}}
\color{green}
\put(235,  120){\makebox(0,0)[l]{\small $1.027\pm0.032$}}
\color{violet}
\put(235,  60){\makebox(0,0)[l]{\small $1.023\pm0.032$}}
\color{gray}
\put(235, 160){\makebox(0,0)[l]{\small $1.002\pm0.005$ (PDG2013)}}
\put(235, 100){\makebox(0,0)[l]{\small $1.065\pm0.046$ (PDG2013)}}
\put(235,  40){\makebox(0,0)[l]{\small $1.067\pm0.047$ (PDG2013)}}
\end{picture}
}
\end{center}
  \caption{Unitarity checks for the first column, second column and second row of the CKM matrix.}
  \label{fig:unitarity_check}
\end{figure}
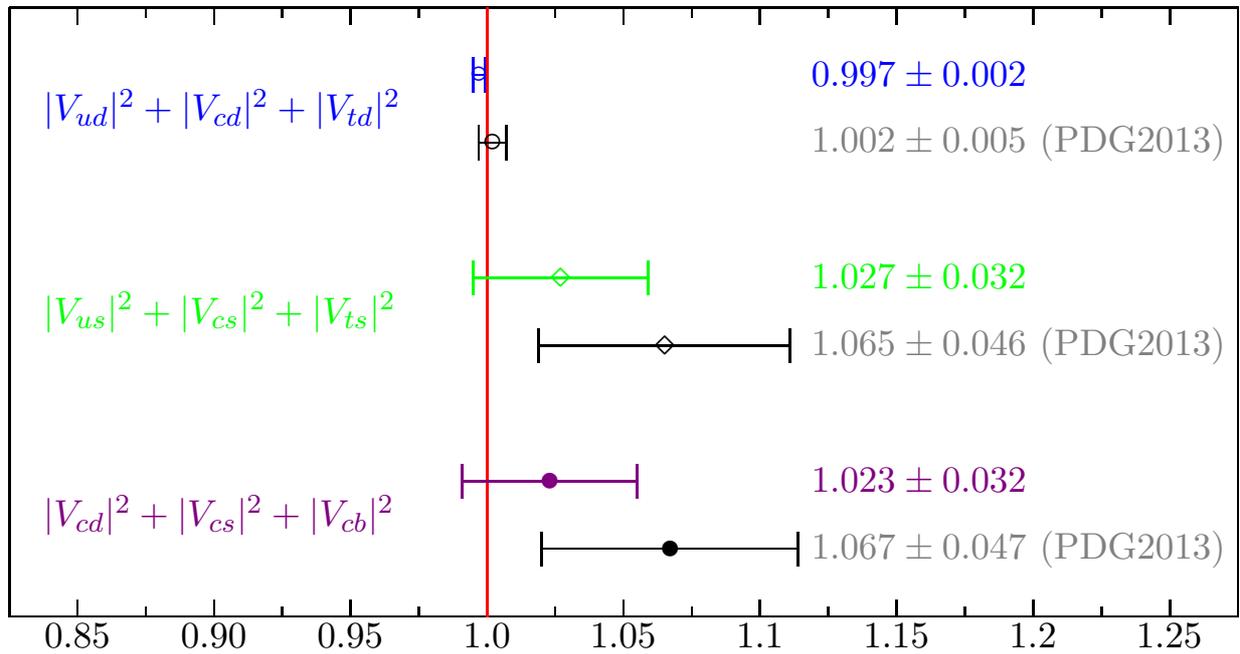

Using the newly extracted $|V_{cd}|=0.218\pm0.005$,
the PDG values $|V_{ud}|=0.97425\pm0.00022$
and $|V_{td}|=(8.4\pm0.6)\times10^{-3}$~\cite{pdg2013},
the first column unitarity of CKM matrix is checked, which is
\begin{equation}\label{eq:U_c1}
    |V_{ud}|^2+|V_{cd}|^2+|V_{td}|^2 = 0.997\pm0.002.
\end{equation}
Using the newly extracted $|V_{cs}|=0.987\pm0.016$,
the PDG values $|V_{us}|=0.2252\pm0.0009$
and $|V_{ts}|=(42.9\pm2.6)\times10^{-3}$~\cite{pdg2013},
we find
\begin{equation}\label{eq:U_c2}
    |V_{us}|^2+|V_{cs}|^2+|V_{ts}|^2 = 1.027\pm0.032
\end{equation}
for the second column of the CKM matrix.
Using these newly extracted $|V_{cd}|$ and $|V_{cs}|$,
and the PDG value $|V_{cb}|=(40.9\pm1.1)\times10^{-3}$~\cite{pdg2013},
we find
\begin{equation}\label{eq:U_r2}
    |V_{cd}|^2+|V_{cs}|^2+|V_{cb}|^2 = 1.023\pm0.032
\end{equation}
for the second row of the CKM matrix.
The unitarity check results for the
first column, second column and second row of the CKM matrix
are shown in Fig.~\ref{fig:unitarity_check} together with the unitarity checks given in PDG2013~\cite{pdg2013}.
The newly determined $|V_{cd}|$ and $|V_{cs}|$ give more stringent checks of the
CKM matrix unitarity compared to those in PDG2013.

\section{Summary}
Combining the precise measurements of leptonic $D^+_{(s)}\to\mu^+\nu_\mu$ decays and semileptonic $D\to\pi(K) e^+\nu_e$ decays
at the CLEO-c, Belle, BaBar and BESIII together with the improved $D^+_{(s)}$ decay constants and semileptonic $D$ decay form factors
calculated in LQCD,
we extract the magnitudes of $V_{cd}$ and $V_{cs}$
to be $|V_{cd}|=0.218\pm0.005$ and $|V_{cs}|=0.987\pm0.016$,
which improve the precisions of those values given in PDG2013 by more than 2.0 and 1.5 factors, respectively.
These improved determinations of $|V_{cd}|$ and $|V_{cs}|$ give more stringent unitarity checks of
the CKM matrix compared to those given in PDG2013.

\section*{Acknowledgements}
This work is supported in part by the Ministry of Science of Technology of China under Contracts No. 2009CB825204; National Natural Science Foundation of China (NSFC) under Contacts No. 10935007 and No. 11305180.

\end{document}